\documentstyle[prl,twocolumn,epsf,floats,aps]{revtex}
\begin{document}
\draft

\twocolumn[\hsize\textwidth\columnwidth\hsize\csname @twocolumnfalse\endcsname

\title{Multi-chain mean-field theory of quasi-one-dimensional quantum 
spin systems}

\author{Anders W. Sandvik}
\address{Department of Physics, University of Illinois at Urbana-Champaign,
1110 West Green Street, Urbana, Illinois 61801 \\
and Center for Nonlinear Studies, Los Alamos National Laboratory,
Los Alamos, New Mexico 87545}
\date{July 12, 1999}

\maketitle

\begin{abstract}
A multi-chain mean-field theory is developed and applied to a two-dimensional
system of weakly coupled $S=1/2$ Heisenberg chains. The environment of a 
chain $C_0$ is modeled by a number of neighbor chains $C_\delta$, $\delta 
= \pm 1,\ldots \pm n$, with the edge chains $C_{\pm n}$ coupled to a 
staggered field. Using a quantum Monte Carlo method, the effective 
$(2n+1)$-chain Hamiltonian is solved self-consistently for $n$ up to $4$.
The results are compared with simulation results for the original Hamiltonian
on large rectangular lattices. Both methods show that the staggered 
magnetization $M$ for small interchain couplings $\alpha$ behaves as 
$M \sim \sqrt{\alpha}$ enhanced by a multiplicative logarithmic
correction.
\end{abstract}


\vskip2mm]

Quasi one-dimensional (1D) quantum spin systems have become an important 
field of study in solid state physics. Many unusual, theoretically predicted 
properties of 1D systems have been observed in real materials. For example, 
the gapless two-spinon spectrum \cite{bethe} of the $S=1/2$ Heisenberg chain 
has been observed in neutron scattering experiments on KCuF$_{\rm 3}$ 
\cite{tennant}, and the Haldane gap predicted for integer $S$ \cite{haldane}
has been detected, e.g., in the $S=1$ compound CsNiCl$_{\rm 3}$ \cite{byers}.
Quantum critical scaling \cite{sachdev} has been observed in the NMR 
relaxation rates of the $S=1/2$ system Sr$_{\rm 2}$CuO$_{\rm 3}$ 
\cite{takigawa1}, possibly even including anticipated \cite{sachdev,starykh}
logarithmic corrections\cite{takigawa2}. In spite of the success of strictly
1D models for these and many other quasi-1D magnetic materials, interchain
couplings can be important as well. A single isotropic chain cannot order,
not even at $T=0$, whereas a transition to a N\'eel ordered state is often 
observed at low temperature; KCuF$_{\rm 3}$ and Sr$_{\rm 2}$CuO$_{\rm 3}$ 
both order at $T_{\rm N} \approx 5$ K. Interchain couplings also change 
qualitatively the nature of the low-lying excitations and lead to interesting
dimensional cross-over phenomena.

One way to take into account interchain couplings $J_\perp$ 
in a quasi-1D system with long-range order is to model the environment 
of a single chain $C_0$ by a staggered magnetic field 
\cite{scalapino,sakai,schulz,prelovsek}. The effective 1D system 
can be solved numerically on small lattices \cite{sakai,prelovsek}, 
or using analytical techniques\cite{schulz}.
In this Letter, the mean-field approach is extended to include 
also a number of neighboring chains $C_\delta$ to which $C_0$ is coupled.
In two dimensions $\delta = \pm 1,\ldots \pm n$. A staggered field is 
coupled to the edge chains $C_{\pm n}$, to model their long-range ordered 
environment. Fluctuations neglected in the environment of $C_\delta$ 
are approximated by a modification of their intrachain interactions, 
in such a way that self-consistency is achieved in the induced staggered 
magnetizations on $C_0$ and $C_\delta$. The effective ($2n+1$)-chain
Hamiltonian can be solved using numerical methods, which typically
perform much better for a few coupled chains than for 2D or 3D
lattices.

Here these ideas will be applied to a system of antiferromagnetic
Heisenberg chains, with the Hamiltonian
\begin{equation}
H = J\sum\limits_{i,j} {\bf S}_{i,j} \cdot {\bf S}_{i+1,j} +
    J_\perp\sum\limits_{i,j} {\bf S}_{i,j} \cdot {\bf S}_{i,j+1},
\label{ham2d}
\end{equation}
where ${\bf S}_{i,j}$ denotes a spin-$1/2$ operator at site $i$ of 
chain $j$. The focus will be on the dependence of the $T=0$ staggered 
magnetization $M=\langle S^z_{i,j} \rangle (-1)^{i+j}$ on the coupling 
constant ratio $\alpha =J_\perp/J$. The question of whether or not 
long-range order $(M > 0)$ develops for arbitrarily small $\alpha > 0$ 
has been the subject of numerous studies. Conventional spin-wave 
theory predicts a finite critical value $\alpha_c$ below which $M=0$ 
\cite{sakai,parola}, but RPA \cite{rosner} gives $\alpha_c = 0$. Some 
self-consistent calculations predict $\alpha_c = 0$ \cite{aoki}, whereas 
others have given $\alpha_c$ as high as $0.2$ \cite{ihle}. Renormalization 
group analyses of the interchain interactions are 
associated with subtleties \cite{affleck1,affleck2}, and completely 
conclusive results have not been presented; however $\alpha_c= 0$ appears
most plausible \cite{affleck2,wang}. An analytical treatment of the
single-chain mean field theory gave the behavior $M \sim \sqrt{\alpha}$
for small $\alpha$ \cite{schulz}. Numerically, $M$ has been 
calculated using exact diagonalization \cite{parola,ihle} and series 
expansion techniques \cite{affleck1}, the former indicating 
$\alpha_c \approx 0.1-0.2$, and the latter giving an upper bound 
$\alpha_c \alt 0.02$. Numerical calculations have in general been 
hampered by convergence problems and difficult extrapolations for 
small $\alpha$. Here multi-chain mean-field calculations will be 
complemented by large-scale quantum Monte Carlo simulations of 
the original 2D Hamiltonian (\ref{ham2d}). It will be shown that quadratic 
lattices are not suitable for extrapolations to the thermodynamic 
limit when $\alpha \ll 1$, due to unusual, non-monotonic finite-size
effects. Using rectangular lattices with aspect ratios $L_x/L_y$ 
as large as $16$ it was, however, possible to study systems with 
$\alpha$ as low as $0.02$. Both the mean-field calculations and the
2D simulations indicate that $M$ vanishes as $\alpha \to 0$ 
{\it slower} than $\sqrt{\alpha}$, due to a logarithmic correction
to this form.

In the conventional single-chain mean-field treatment of the Hamiltonian
(\ref{ham2d}) \cite{scalapino,sakai,schulz}, the coupling of a chain
$j$ to its nearest-neighbors $j \pm 1$ is approximated by 
$J_\perp  \sum_{i}S^z_{i,j} [\langle S^z_{i,j-1} \rangle + \langle 
S^z_{i,j+1} \rangle ]$. In a N\'eel state $\langle S^z_{i,j} \rangle 
= (-1)^{i+j}M$, and one obtains an effective 1D Hamiltonian,
\begin{equation}
H_1 = J\sum\limits_{i} {\bf S}_{i} \cdot {\bf S}_{i+1} -
     h\sum\limits_{i} (-1)^i S^z_{i},
\label{hstagg}
\end{equation}
with the self-consistency condition $h=2J_\perp M$ which directly relates
$M(h)$ to $M(J_\perp)$ of the 2D system. 

The idea of the multi-chain
mean-field theory is to model the environment of a chain $C_0$ by its
first few neighbor chains $C_\delta$, $\delta = \pm 1,\ldots \pm n$,
with only the edge chains  $C_{\pm n}$ coupled to a staggered field. This
induces a staggered magnetization in all chains. The dynamic environment 
for $C_0$ provided by the $C_\delta$ chains should be considerably 
more realistic than just the static staggered field of the single-chain 
theory. If $C_0$ and $C_\delta$ are identical chains, it is not possible 
to obtain a self-consistent description, however. The staggered magnetization
will be largest at the edges and decrease towards the center, due to the 
neglected quantum fluctuations at the edges. These fluctuations can be 
approximated by a modification of the intrachain interactions of $C_\delta$.
There are clearly many possible ways of doing this, and the optimum way, that
would best mimic the presence of an infinite half-plane of other chains, is 
not obvious. One requirement is that the additional interactions have to be 
invariant under spin rotations in the $xy$-plane (since the field breaks the 
$O(3)$ symmetry, an $O(2)$ symmetric effective interaction in $C_\delta$
is permissible). Here the simplest interaction satisfying this requirement will
be considered, namely, the $xy$ part of the coupling is given a strength 
$J^{xy}_{|\delta|} = J(1 + \lambda_{|\delta|})$ different from $J^z_{|\delta|}
= J$. Increasing $\lambda_{|\delta|} > 0$ increases the quantum fluctuations.
The $(2n+1)$-chain effective Hamiltonian is then
\begin{eqnarray}
H_n = &J&\sum\limits_{i=1}^L \sum\limits_{j=-n}^n 
{\bf S}_{i,j} \cdot {\bf S}_{i+1,j} + 
J_\perp\sum\limits_{i=1}^L\sum\limits_{j=-n}^{n-1} 
{\bf S}_{i,j} \cdot {\bf S}_{i,j+1} 
\nonumber \\
& + & \sum\limits_{i=1}^L\sum\limits_{|\delta|=1}^n  \lambda_{|\delta|} 
(S^x_{i,\delta} S^x_{i+1,\delta} + S^y_{i,\delta} S^y_{i+1,\delta}) 
\label{hmulti} \\
& + & h\sum\limits_{i=1}^L (-1)^i (S^z_{i,-n} + S^z_{i,n}). \nonumber
\end{eqnarray}
There are $n+1$ self-consistency conditions:
\begin{eqnarray}
M & \equiv & M_0 = M_1 = ,\ldots ,=M_n , \nonumber \\
h & = & MJ_\perp .
\label{conditions}
\end{eqnarray}
\label{selfconst}
Since the environment of a chain $C_k$ becomes more similar to that
of the real 2D system the closer it is to the center $(k=0)$ of the effective
$2n+1$ chain system, the self-consistent parameters can be expected to
satisfy $0 < \lambda_1 < \ldots < \lambda_n$. For a given $\alpha$,
the magnetization (as well as other properties) should converge to 
its correct value as $n,L \to \infty$. Therefore, the details of the 
intrachain interactions used to achieve self-consistency can be 
seen to be unimportant; they will only affect the rate of convergence
with increasing $n$. 

Here quantum Monte Carlo results for the cases $n=1,2,3$, and $4$ will 
be presented. In addition, results will be shown for the conventional 
single-chain effective Hamiltonian (\ref{hstagg}), which 
corresponds to $n=0$. This 1D Hamiltonian has been studied by Schulz 
via a mapping to a solvable fermion model in the continuum limit 
\cite{schulz}, with the result $M =0.719 \sqrt{\alpha}$ for the 2D 
system. The mapping has not been demonstrated rigorously, however, and 
the above form of $M$ can furthermore be valid only for small $\alpha$. 
Numerical calculations have previously been carried out for $L \le 10$
\cite{sakai}, which is not sufficient for addressing the 
behavior for $\alpha \ll 1$ in the thermodynamic limit.

\begin{figure}
\centering
\epsfxsize=7.5cm
\leavevmode
\epsffile{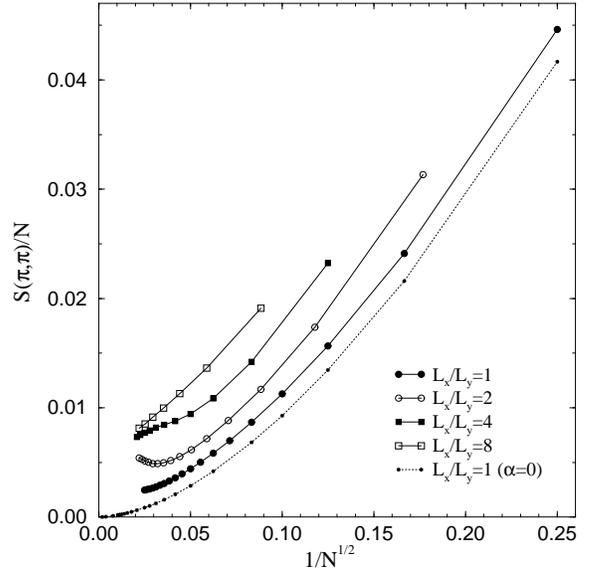}
\vskip1mm
\caption{Quantum Monte Carlo results for the staggered structure factor
on rectangular lattices with different aspect ratios and $\alpha=0.05$. 
The behavior for a quadratic lattice with $\alpha=0$ (independent 1D
chains) is also shown. Statistical errors are much smaller than the
symbols.}
\label{fig1}
\end{figure}

Before presenting the mean-field results, quantum Monte Carlo calculations 
for the full 2D Hamiltonian (\ref{ham2d}) will be discussed. For the spatially 
isotropic system ($\alpha=1$), very accurate results for $M$ have previously 
\cite{num2d} been obtained using ground state results for the staggered 
structure factor, 
\begin{equation}
S(\pi,\pi)={1\over N} \sum\limits_{i=1}^N\sum\limits_{j=1}^N
\langle S^z_iS^z_j \rangle (-1)^{(x_i-x_j+y_i-y_j)}.
\end{equation}
Accounting for rotational averaging, the sublattice magnetization is given by
\begin{equation}
M^2 = 3S(\pi,\pi)/N  ~~~~~ (N \to \infty).
\label{m2}
\end{equation}
Here this quantity will be extrapolated to infinite size for $\alpha < 1$. 
Using the stochastic series expansion method with an efficient cluster update
\cite{sse}, ground state properties of systems with several thousand spins
can be studied (for the largest lattices considered here, inverse temperatures 
$\beta=J/T$ as high as $2048$  were used in order to obtain results free of 
temperature effects). 

For $\alpha=1$, the leading finite-size corrections to $M^2$ as 
defined in Eq.~(\ref{m2}) are positive and $\sim 1/\sqrt{N}$ \cite{num2d}. 
This can be expected also for $0 < \alpha < 1$ if the system is ordered. 
Fig.~\ref{fig1} shows results for $\alpha=0.05$ on $L \times L$ lattices 
with $L$ up to $40$. The results extrapolate to $M > 0$, but subleading 
corrections to the linear behavior are clearly large. Previously, results 
for smaller $L$ were used as evidence that $M$ vanishes below a critical
value $\alpha_c \sim 0.1 - 0.2$ \cite{parola,ihle}. Results for rectangular
lattices with different aspect ratios $R=L_x/L_y$ reveal a considerable 
dependence on $R$, as also shown in Fig.~\ref{fig1}. For $R=8$, the expected
linear behavior can be seen clearly, and for $R=4$ there is a cross-over 
to this behavior for large systems. For $R=2$ there is a clear minimum, 
and the $R=1$ results also suggest one. In the two latter cases the 
finite-size behavior is hence non-monotonic, and there has to be a maximum
for even larger systems before the asymptotic, linear (with positive slope) 
approach to the infinite-size value, which for $\alpha=0.05$ is
$S(\pi,\pi)/N \approx 0.0056$ (from an extrapolation of the $R=8$ data).

The non-monotonicity can be understood as resulting from a cross-over from 
1D to 2D behavior. A chain of length $L_x$ has an excitation gap $\Delta(L_x)
\sim 1/L_x$. If this gap is larger than the effective energy scale of the 
coupling of the chains, i.e., the spin-stiffness $\rho_s^y$, then the system 
essentially behaves as a system of 1D chains, with exponentially 
damped correlations between the chains. A cross-over to 2D behavior can be 
expected when $\Delta(L_x) \sim \rho_s^y$, which occurs for smaller system 
sizes $N=L_xL_y$ when the aspect ratio $R$ is large, in agreement with the 
results in Fig.~\ref{fig1}. When $\alpha \ll 1$ (and therefore 
$\rho_s^y \ll 1$), quadratic lattices therefore have to be very large for 
extrapolations to infinite size to be meaningful. Instead, rectangular 
lattices with $R$ increasing with decreasing $\alpha$ should be used. 
Using aspect ratios as large as $R=16$, the sublattice magnetization was 
calculated for $\alpha$ as small as $0.02$. Below, the results will be 
compared with the single- and multi-chain mean field theories. 

In the single-chain theory the magnetization curve $M(J_\perp)$ is 
directly obtained from a calculation of $M(h)$ for the Hamiltonian 
(\ref{hstagg}). The effective model (\ref{hmulti}) depends explicitly
on $J_\perp$, however, and for each $J_\perp$ a search for the 
self-consistent values $h,\lambda_1,\ldots,\lambda_n$ is required. With 
the Monte Carlo method used \cite{sse}, 
the derivatives  $\partial M_j/\partial h$
and $\partial M_j / \partial \lambda_k$ can also be calculated. Using 
these, an iterative scheme where
\begin{eqnarray}
h(m+1) & = & h(m) + \Delta_h(m), \nonumber \\
\lambda_k(m+1) & = & \lambda_k (m) + \Delta_{\lambda_k}(m),
\end{eqnarray}
can be employed, starting from estimated values $h(0)$ and 
$\lambda_k(0)$. The self-consistency conditions (\ref{conditions}) give 
the corrections $\Delta_h(m)$ and $\Delta_{\lambda_k}(m)$ as the 
solution of $n+1$ coupled equations, e.g., for $n=1$,
\begin{eqnarray}
\Delta_h & & 
(\partial M_0 / \partial h - \partial M_1/ \partial h ) \nonumber \\
& & +\Delta_{\lambda_1}
(\partial M_0 / \partial \lambda_1  - \partial M_1 / \partial\lambda_1 ) =
M_1 -  M_0 , \\
\Delta_h & &
(\partial M_0 / \partial h + \partial M_1 / \partial h - 2/J_\perp )
\nonumber \\ 
& & + \Delta_{\lambda_1} 
(\partial M_0 / \partial \lambda_1 + \partial M_1 / \partial\lambda_1 ) = 
2h/ J_\perp - M_1 - M_0 . \nonumber
\end{eqnarray}
Self-consistency is typically achieved this way in as few as two or 
three iterations. 

For a finite system, the self-consistent $M$ vanishes
below a critical value $\alpha_c (L)$ which decreases with increasing
$L$. In order to study the behavior for small $\alpha$ very large $L$ 
have to be used. The largest sizes used here were $L=1024$ 
for $n=0$, $512$ for $n=1,2$, and $256$ for $n=3,4$. Inverse
temperatures $\beta=J/T$ as high as $2L$ were used in
order to completely project out the ground state.

\begin{figure}
\centering
\epsfxsize=7.5cm
\leavevmode
\epsffile{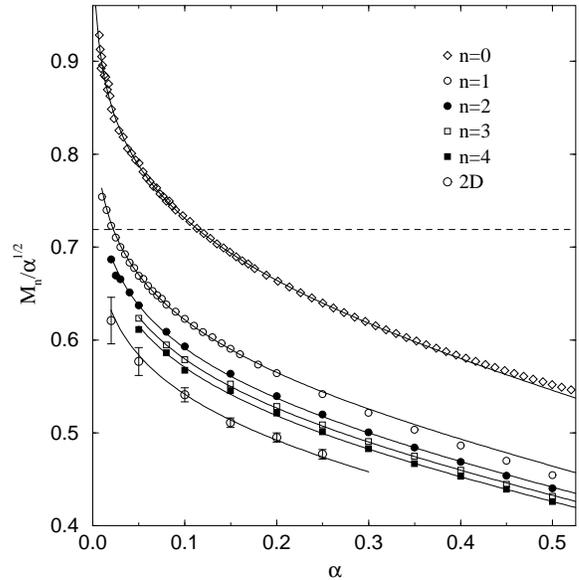}
\vskip1mm
\caption{Self-consistent staggered magnetization vs interchain 
coupling in the single-chain mean-field theory ($n=0$) and multi-chain 
mean-field theories with $n=1-4$. Statistical errors are at most comparable 
to the symbol sizes. Monte Carlo results for the full 2D Hamiltonian
are shown with estimated error bars. The dashed line is the analytical 
$n=0$ result [11]. The solid curves are of the form $M_n/\sqrt{\alpha}
= A_n(1+b\alpha) {\rm ln}^\gamma (a/\alpha)$, with $b=0.095$, $a=1.3$, 
and $\gamma =1/3$ in all cases. These parameters, and $A_0=0.529$, were 
chosen to fit the $n=0$ data. Only the amplitudes $A_n$ were subsequently 
adjusted to fit the other data sets.}
\label{fig2}
\end{figure}

\begin{figure}
\centering
\epsfxsize=7.5cm
\leavevmode
\epsffile{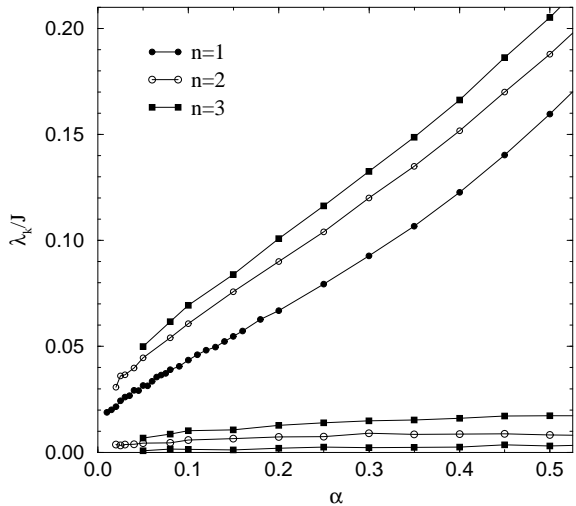}
\vskip1mm
\caption{Self-consistent anisotropy parameters vs interchain coupling 
for $n=1,2,3$. All $\lambda_k$, $k=1,\ldots ,n$, for given $n$ are shown
using the same symbols, and in all cases $\lambda_1 <\ldots <\lambda_n$.}
\label{fig3}
\end{figure}

All results for $M$, including those for the original 2D Hamiltonian 
(\ref{ham2d}) extrapolated to infinite size, are shown divided by 
$\sqrt{\alpha}$ in Fig.~\ref{fig2}. The behavior predicted by Schulz 
\cite{schulz} using a mapping of the $n=0$ mean-field theory to a 
solvable continuum model should then be a constant. The numerical 
results for $n=0$ do not agree with this; instead $M_0/\sqrt{\alpha}$ 
appears to diverge as $\alpha \to 0$. The behavior for $\alpha \alt
0.4$ is closely reproduced by the form $M_0 = A_0\sqrt{\alpha}(1+b\alpha)
{\rm ln}^\gamma (a/\alpha)$, with $\gamma \approx 1/3$, $A_0\approx 
0.53$, $a \approx 1.3$, and $b \approx 0.1$. 
This result shows that the mapping of 
Eq.~(\ref{hstagg}) to the  continuum model is not exact. A reason for this 
could be the presence of marginally irrelevant operators, which are known 
to lead to logarithmic corrections to physical observables in the case $h=0$ 
\cite{logs}. The results for higher $n$ also show a similar divergent 
behavior, but with the available computer resources it was not possible 
to extend the calculations to as small $\alpha$ as for $n=0$. The above
logarithmic form fits quite well also all the multi-chain results, with
{\it only} the over-all factors $A_n$ adjusted. This is a strong indication
that the logarithmic correction survives in the 2D limit ($n \to \infty$). The 
curves indeed approach the results obtained using finite-size extrapolations
for rectangular 2D lattices, confirming that the multi-chain mean-field theory
converges correctly. Remarkably, the same expression that describes all 
the mean-field data also fits the 2D results, with the amplitude 
$A_{2D} \approx 0.39$. 

The self-consistent values of the $xy$-anisotropy parameters
are graphed in Fig.~\ref{fig3} for $n=1,2,3$. For $n > 1$, the anisotropy
is always largest at the edges, as expected, and rapidly decreases 
as the center chain is approached. The behavior for $\alpha \to 0$ 
suggests a very slow asymptotic decay to zero --- again an indication 
of log-corrections.

To conclude, both the multi-chain mean-field theory and calculations for 
the original Hamiltonian strongly support a critical coupling $\alpha_c=0$,
and a staggered magnetization that for small interchain couplings behaves 
as  $M \sim \sqrt{\alpha}$ enhanced by a logarithmic correction. In 
the conventional single-chain mean-field theory ($n=0$), all 
interchain quantum fluctuations are neglected. 2D quantum fluctuations 
develop systematically in the multi-chain theory as $n$ is increased.
For $\alpha \ll 1$, the functional form of the sublattice magnetization 
is the same for all $n$ considered ($n=0-4$), indicating that the interchain 
quantum fluctuations only affect the over-all magnitude of $M$. 
Hence even the conventional single-chain theory gives the correct 
functional form for $M$, although the magnitude is over-estimated by a 
factor $\approx 1.35$. The previous analytical treatment of the 
single-chain theory \cite{schulz} misses the log-correction.

For the model considered here, it was possible to explicitly test
the multi-chain mean-field theory against large-scale quantum Monte
Carlo results. In general this would not be possible, e.g., for
systems with frustrated interactions where Monte Carlo simulations
suffer from sign problems. The density matrix renormalization group 
method \cite{dmrg} could be used to study the effective multi-chain 
models in such cases.

I would like to thank D. K. Campbell and H. J. Schulz for stimulating 
discussions. This work was supported by the NSF under grant 
No.~DMR-97-12765. Some of the numerical calculations were carried out 
at the NCSA.

\end{document}